\documentstyle[floats,multicol,aps,epsf,prb]{revtex}

\newcommand\ltdash{\raise-1.8pt\hbox{$\scriptscriptstyle |$}}
\newcommand \beq  {\begin{equation}}
\newcommand \eeq  {\end{equation}}
\newcommand \bea {\begin{eqnarray} }
\newcommand \eea {\end{eqnarray}}

\newcommand \uru {$URu_2Si_2$}

\begin{document}
\draft
\twocolumn[\hsize\textwidth\columnwidth\hsize\csname @twocolumnfalse\endcsname
\title{
Hidden Order in $URu_2Si_2$
}
\author{ N. Shah,$^{1}$ P. Chandra,$^{2}$
P.Coleman$^{1}$ and  J.A. Mydosh$^{1-4}$}
\address{$^1$Department of Physics and Astronomy,
Rutgers University,
136 Frelinghausen Road,
Piscataway, NJ 08854-8019, USA
}
\address{$^2$NEC Research, 4 Independence Way,
Princeton, NJ
}
\address{$^3$Lucent Bell Laboratories,
100 Mountain Ave, Murray Hill, NJ 07904, USA
}\address{$^4$ 
Kamerlingh Onnes Laboratory, 
Leiden University,
PO Box 9506, 2300 RA Leiden,
The Netherlands }
\maketitle
\date{\today}
\maketitle
\begin{abstract}
We review current attempts to
characterize the underlying nature of the
hidden order in $URu_2Si_2$. 
A wide variety of experiments point to the
existence of two order parameters: a large
primary order parameter of unknown character
which co-exists with secondary antiferromagnetic
order. Current theories can
be divided into two groups determined by whether or not
the primary order parameter breaks time-reversal symmetry.
We propose a series of experiments designed
to test the time-reversal nature of the underlying
primary order in $URu_2Si_2$  and to characterize
its local single-ion physics.

\end{abstract}
\vskip 0.2 truein
\newpage
\vskip2pc]


The nature of the hidden order parameter in $URu_2Si_2$ is a longstanding mystery
in heavy fermion physics.\cite{buyers96}
At 17 K this material undergoes a second-order phase transition characterized by
sharp features in bulk properties including specific
heat,\cite{palstra85} linear\cite{palstra85,maple86}
and nonlinear\cite{miyako91,ramirez92} susceptibilities, thermal expansion\cite{devisser86} and
resistivity.\cite{palstra87} The accompanying gap in the magnetic excitation
spectrum,\cite{walter86,mason91} also indicated by the
exponential dependence of the specific heat below the transition
$\Delta C_V \propto e^{- \Delta/T} $,
suggests the formation of an itinerant
spin density wave at this temperature. 
However, the size of
the observed staggered moment\cite{broholm87} ($m_0 = 0.03\mu_B$) {\sl cannot}
account for the bulk properties, e.g. the entropy
loss and the size of the gap which develops
at the transition.  This mismatch between the tiny ordered moment and the large
entropy of condensation indicates the presence of a primary order parameter
whose nature remains to be characterized.

Two sets of recent developments provide impetus
for a renewed discussion of this material.  In particular high-field measurements
have emphasized the distinction between the hidden primary and
the secondary magnetic order parameters. Though measurements of the high-field
resistance,\cite{mentink96} thermal expansion\cite{mentink97} and
specific heat\cite{vandijk97} indicate
that the primary order parameter is
destroyed by a field of $40$ Tesla, neutron scattering results suggest that the
magnetic order may disappear at a much lower field strengths.\cite{mason95}
On a separate front, measurements
of the specific heat, susceptibility and thermal expansion\cite{amitsuka93,amitsuka97}
on dilute $U$ in $Th_{1-x}U_xRu_2Si_2$ have
provided new insight into the uranium single-ion physics of this family of
materials. Both of these
quantities display a logarithmic dependence on temperature that is
suppressed by a magnetic field, features suggesting the presence
of a non-Kramers, $\Gamma_5$ {\sl  magnetic} doublet. Unlike a Kramer's
doublet, this ionic ground-state 
can be split by {\sl both} magnetic and strain fields.  These two new sets of
observations motivate us to propose further experiments designed
to distinguish between various characterizations of the hidden order.

Many competing theories have been proposed for the primary hidden
order in \uru.  The emphasis of these theoretical proposals
has been on the microscopic order parameter. Broadly speaking, these theories
divide into two distinct categories.
In the first set, from here onwards designated as (A),
the primary order parameter {\sl breaks} time-reversal symmetry;
proposals include
three-spin order,\cite{gorkov93} spin density waves in higher angular momentum
channels,\cite{ramirez92} and antiferromagnetic states with strongly renormalized
g-factors.\cite{sikema96,ikeda98} 
By contrast the primary order parameter in category (B)
is {\sl invariant} under time-reversal symmetry,
and staggered quadrupolar order\cite{santini98} and Jahn-Teller
distortions\cite{kasuya97} are examples in this classification scheme.
Unfortunately experiment has 
been unable to clearly distinguish between these different microscopic
proposals.  

In this paper, we should like to turn the debate in a
more phenomenological direction. We argue that as a necessary prelude
to the development of a theory for the microscopic order parameter
in \uru, we need to ask two key questions:
\begin{itemize}

\item Does the primary order parameter break time reversal symmetry?

\item What single-ion physics governs the low energy behavior
in  stoichiometric
\uru?

\end{itemize}
At present neither question has been definitively answered, and to this
end we propose a set of experiments designed to address these issues.

The ideal framework for our phenomenological discussion about the
order parameter is Landau-Ginzburg theory.  In this context, the distinction
between theories in categories (A) and (B) lies in the allowed couplings
between the primary and the secondary order parameters.
Let us denote the primary and secondary 
order parameters  by $\psi$ and $m$, respectively.  Quite generally
the Landau-Ginzburg free energy must contain three terms
\bea
{\cal F}[\psi,m] = {\cal F}_1[\psi] + {\cal F}_2[m] + {\cal F}_c[\psi,m]
\eea
A number of experiments suggest that the hidden order
is staggered.\cite{ramirez92} Uniform order parameters tend to couple directly to macroscopic
properties, e.g. the uniform magnetization and
thus cannot be easily hidden. 
Under this assumption, the free energy must satisfy
\bea
{\cal F}[\psi,m] 
={\cal F}[-\psi,-m].
\eea 
Since antiferromagnetism seems to develop simultaneously with
the hidden order, it is natural to consider coupling terms of 
the form
\bea
{\cal F}^{(A)}_c(\psi,m)= g_A m \psi.
\label{couplingA}
\eea
As magnetization breaks time-reversal symmetry,
and is of odd parity under time-reveral, such 
a term is only permitted if $\psi$ is also odd under time reversal,
and thus also breaks time-reversal
symmetry. Such terms can only occur in models of type (A)
where $\psi$ {\sl breaks} time-reversal symmetry.
In theories of type (B) where $\psi$
is even under time-reversal invariance
the simplest coupling
consistent with both time-reversal symmetry and translational invariance
takes the form
\bea
{\cal F}^{(B)}_c(\psi,m)= g_B m^2 \psi^2
\label{couplingB}
\eea 
Note that terms of the form $m^2 \psi$ and $m \psi^2$ are ruled out
if the hidden order is staggered and is invariant
under time-reversal symmetry.\cite{walker95}
These two types of coupling, (\ref{couplingA}) and
(\ref{couplingB}),
lead to very different
predictions for the $H-T$ phase diagram.

In order to understand these distinctions, let us write
the separate free energies for the secondary and primary order parameters.
For both categories of theory, the primary free energy takes the form
\bea
{\cal F}_1[\psi]= -\alpha t \psi^2 + \beta \psi^4 + \alpha h^2 \psi^2
\eea
where 
$t= (T_c-T)/T_c$ is the reduced temperature, 
measuring  the deviation from the 
transition temperature $T_c$ of the primary order parameter
and $h= H/H_c$ is the ratio between the external magnetic field
and the measured critical field at zero temperature ($H_c= 40 T$).
Translational
invariance is enough to rule out a linear
coupling between
$h$ and $\psi$ in both categories of theory.  This form of the
free energy is broadly consistent with many of the observed phenomenon.
We can rewrite ${\cal F}_1$ in the form
\bea
{\cal F}_1[\psi] = \beta\left[ \psi^2 - \psi_0^2(h,t)\right]^2 + F_1
\eea
where
\bea
\psi_0(h,t) = \left[ \frac{\alpha}{2 \beta} (t-h^2)\right]^{\frac{1}{2}},
\eea
is the equilibrium value of the primary order parameter, and
\bea
F_1= - \beta \psi_0^4(h,t)= - \frac{\alpha^2}{4 \beta} (t-h^2)^2
\eea
is the equilibrium free energy.

If we ignore the coupling to the secondary order parameter, then by reading
off the various derivatives with respect to temperature and field, 
we are able to deduce that
\bea
\Delta \left(\frac{C_v}{T}\right) &=& - \frac{1}{T_c^2}\frac{\partial^2 F_1}{\partial t^2} =
\frac{1}{2T_c^2}\zeta \cr
\Delta \left(  \frac{d \chi}{dT}\right) &=& 
- \frac{1}{H_c^2 T_c}\frac{\partial^3 F_1}{\partial t \partial h^2}
=
- \frac{1}{H_c^2 T_c}\zeta
\cr
\Delta \chi_3 &=& 
- \frac{1}{H_c^4 }\frac{\partial^4 F_1}{\partial h^4}=
\frac{6}{H_c^4}\zeta
\eea
where $\chi_3 = - \partial^4 {\cal F}_1/\partial H^4$ is the non-linear
susceptibility and we have denoted $\zeta = \alpha ^2/\beta$.
From these three results, we can deduce the relationship
\bea
\Delta \left(\frac{C_v}{T} \right)
 \Delta \chi_3
= 3 \left[\Delta\left(\frac{d \chi}{dT}\right) \right]^2  
\eea 
This result
is in good accord with the measured anomalies in this
material.\cite{chandra94}
This agreement indicates 
that the phase transition is well described by mean-field theory,
though it does not reveal any specifics about the nature of
the hidden order.
As an aside, we note that if the transition were associated with
a conventional spin density wave, this expression would
become
\bea
\Delta \left(\frac{C_v}{T} \right)
 \Delta \chi_3
= 3\left(m_0^4\right)
\eea 
where $m_0$ is the staggered moment; this relation
is clearly {\sl not} obeyed\cite{ramirez92} in $URu_2Si_2$ where the
anomalies in the specific heat and the nonlinear
susceptibility are large and $m_0 = 0.03 \mu_B$.  Thus
a conventional spin density wave can be excluded as
a possible cause for the hidden order in $URu_2Si_2$.

Let us now consider the  way in which theories of type
(A) and (B) differ. In type (A) theories, the
quartic terms in ${\cal F}_2$ may be neglected,
and it is sufficient to take
\bea
{\cal F}_2^{(A)}[m] = a(h) m ^2 + O(m^4)
\eea
where $a(h)$ is positive. 
Now since a magnetic field
always raises the energy of an antiferromagnet,
we may write
\bea
a(h) = a[1 + \delta h^2].\label{mag2}
\eea 
This means that at reduced 
fields above the scale 
$h\sim 1 /\sqrt{\delta}$
($H\sim H_c/\sqrt{\delta}$),  
the energy of 
the induced order parameter ${\cal F}_2^{(A)}= a(h)m^2$ is dominated by its coupling to the
external magnetic field. 
Furthermore in these theories in category (A) where
$\psi$ breaks time-reversal symmetry, the
linear coupling between the primary and secondary order parameters
means that the primary order parameter always induces a staggered
magnetic moment.  If we assume $g_A$ is small, then by minimizing
${\cal F} = {\cal F}_1 + {\cal F}_2 +{\cal F}_c^A$, we obtain
\bea
m = - \frac{g_A}{2 a(h) } \psi_0(h,t)\label{mag1}
\eea
The small magnitude of the magnetic order parameter in scenario (A)
arises naturally from the assumed small magnitude of $g_A$.
From the high field
experiments,\cite{mason95,mentink96,mentink97,vandijk97} 
it is known that the 
field-dependence of $m$ at low temperatures is more rapid than that
of the primary order parameter. 
Using (\ref{mag2}) and (\ref{mag1}), at $T=0$, 
\bea
m[h]= m_0\frac
{[ 1- h^2]^{\frac{1}{2}}}{1+ \delta h^2}\eea
where $m_0 = - \frac{g_A}{2a}\sqrt{\frac{\alpha}{2\beta}}$.  
We see that the staggered magnetization is then a product of a Lorentzian
times the field dependence of the hidden order parameter,$\psi$. 
For small fields
\bea
m = m_o \left[ 1 - \frac{1}{2}\left(\frac{h}{h_m}^2\right) \right]
\eea
where 
\bea
h_m = \frac{1}{\sqrt{1 + 2\delta}}
\eea
$h_m$ sets the magnitude of the field scale
where the secondary order
vanishes, based on a low field extrapolation of the magnetization.
Since the magnetization is always finite for
$\psi\ne 0$, scenario (A) necessarily implies that there
will be a point of inflection in the field dependence of the staggered
magnetization around the field value $H_m \sim H_c h_m$;  
at field strengths greater than $H_m$, the energy of the staggered order
is dominated by its coupling to the external magnetic field, but
the staggered order is prevented from going to zero by its
coupling to the hidden primary order. 
In Fig. 1 we show a typical curve for $m(h)/m_0$. 
The absence/presence of a point of inflection in $m(h)$ is a key
experimental
test for
scenario (A).
\begin{figure}[btp]
\epsfxsize=3.0truein \epsfbox{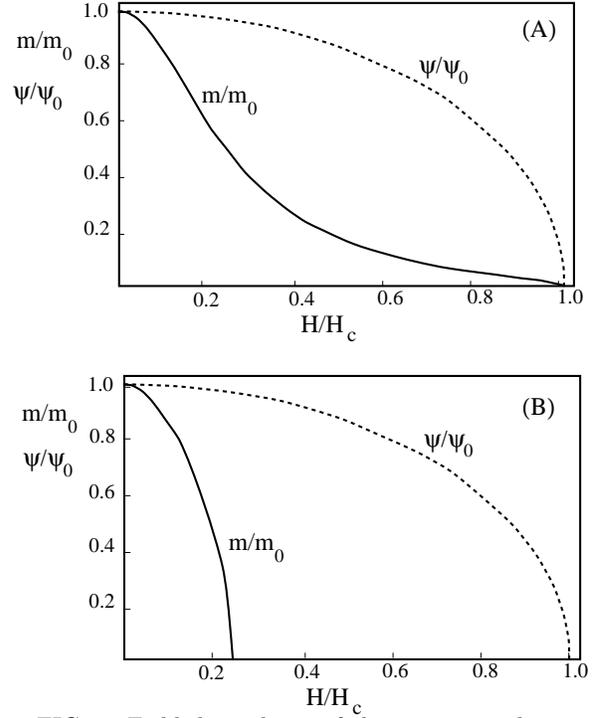}
\protect\caption{Field dependence of the primary
order parameter and the staggered magnetization
in scenarios  (A) and (B). When the primary order parameter
breaks time reversal symmetry, the staggered magnetic order remains
finite so long as the primary order is present, leading to a point
of inflection in the magnetization versus field. 
}
\label{fig1}
\end{figure}

Let us now turn to scenario (B). In this case, it is necessary to
assume that the system is close to an antiferromagnetic instability,
so that
\bea
{\cal F}_2^{(B)} +{\cal F}_c^{(B)} = 
a (T_m- T) m^2 + b m^4 + g_B m^2 \psi^2
\eea
We can rewrite this in the form
\bea
{\cal F}_2^{(B)} +{\cal F}_c^{(B)} = 
a (T_m[\psi]- T) m^2 + b m^4 
\eea
where $T_m[\psi]= T_m-\frac{g_B}{a}\psi^2$.  
Clearly at temperatures close to $T_c$
where  $\psi$ is small the renormalization
of $T_m$ is negligible,
so
that the coupling between the two order parameters
can be effectively neglected.
Experimentally the transitions
associated with the development of the primary and secondary
order parameters are identical to within the accuracy of
the measurements;\cite{mason95,mentink96,mentink97,vandijk97}
thus to fit these observations proponents of
model (B) must set
\bea
T_m= T_c
\eea
This relies on coincidence, since within
scenario (B) the coupling between the order parameters
does not contribute towards linking the two transitions
and they are 
therefore truly independent.
We note that staggered quadrupolar order is an example
of a primary order parameter in class (B); all
known systems with {\sl continuous} double quadrupolar-magnetic
transitions have a separation in the two
temperature scales.\cite{morin90}
If the primary phase transition of $URu_2Si_2$ had been
{\sl discontinuous} (e.g. first-order) then this requirement ($T_m = T_c$) could have
been relaxed, and indeed there is such an example
of a first-order quadrupolar-magnetic transition\cite{becker97} 
in $U_2Rh_3Si_5$.
However field-dependent measurements in $URu_2Si_2$ clearly indicate that
the primary order parameter grows continuously as the temperature
is reduced, ruling out this
possibility.\cite{mason95,mentink96,mentink97,vandijk97} 

A second aspect of scenario (B) concerns the size of the staggered
magnetization. In order to account for the small size of the staggered moment,
we require that
\bea
m_o=
\sqrt{
\frac{ a T_m}{2b}
}
\eea
is naturally small. A microscopic theory would have to account for
the magnitude of this parameter.  It is worth noting that
in the related  heavy fermion superconductors, $UPd_2Al_3$ and
$UNi_2Al_3$, which also exhibit co-existent magnetism and superconductivity,
the size of the ordered moment is large.
In scenario (B) the field-dependence of the staggered order parameter
is then entirely independent of the primary order parameter.

In Fig. 2 we contrast the phase diagrams
expected in the two different scenarios (A) and (B). 
The qualitative distinction
is quite striking and immediately suggests a ``tie-breaking'' experiment.
If the underlying order parameter is indeed of type (A), then high-field neutron scattering
experiments should observe a marked inflection in the field-dependence of the staggered
magnetization; this should occur
long before the upper critical field ($H \approx 40T$) of the primary order
parameter is reached.
\begin{figure}[btp]
\epsfxsize=3.0truein \epsfbox{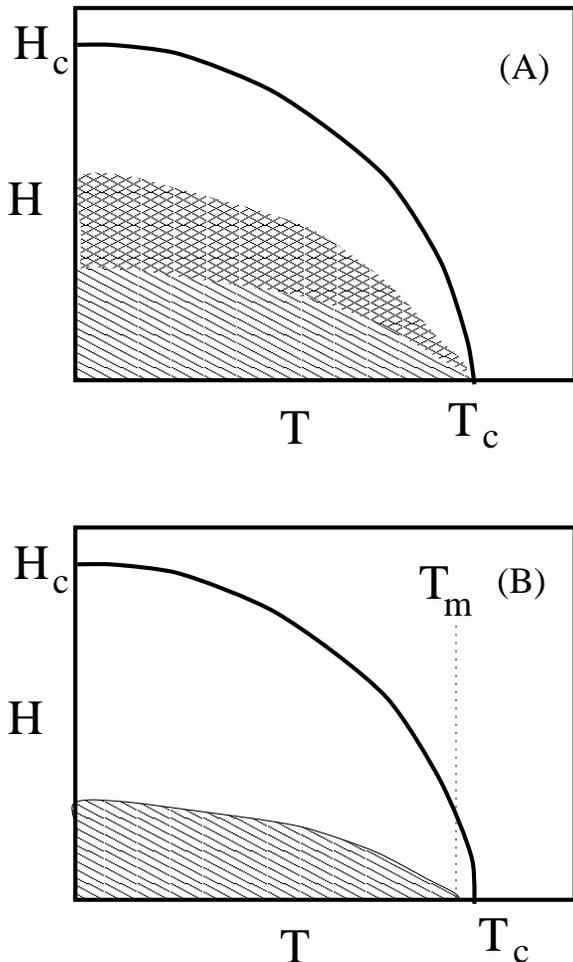}
\protect\caption{The contrasting phase diagrams
for 
scenarios  (A) and (B). In (A), where the primary
order parameter has broken time-reversal symmetry,
the staggered magnetic order remains
finite so long as the primary order is present. The cross-hatched
area refers to where $m(h,t)$ has a region of inflection
(see Figure 1).
In (B),
there is a sharp phase transition at a finite field and 
$T_m$ and $T_c$ match up only by coincidence. 
}
\label{fig2}
\end{figure}

We now turn to the second part of our discussion and consider the nature
of the single ion physics.  Any microscopic theory is critically
dependent on this physics. 
For example,  Santini\cite{santini98} has
proposed that the key physics of $URu_2Si_2$ is governed by the mixing of two
non-degenerate singlet ground-states leading to a staggered quadrupolar
ground-state.
On the other hand, Amitsuka et al.\cite{amitsuka93} suggest that a different
ground-state is relevant to dilute concentrations of Uranium in 
$ThRu_2Si_2$, involving a
magnetic non-Kramers doublet of a type first considered by Cox and Makivic.\cite{cox94}  Were such a ground-state to survive to the dense
system, it would lead to a magnetic two-channel Kondo lattice.
This immediately suggests three distinguishing experiments:\begin{itemize}

\item  A definitive test of the proposal by Amitsuka et al.\cite{amitsuka93,amitsuka97} for dilute $U$
concentrations has not yet been performed.  Theory predicts
that if the ground-state is that of a two-channel Kondo model,
then at finite magnetic fields the logarthmic divergence of
$\gamma = C_v(T)/T$ will be cut-off by a Schottky anomaly with an associated entropy of
$\frac{1}{2}ln 2$. This fractional entropy is distinctive of the
two-channel Kondo model and heuristically arises from the partial
quenching of the fermionic degrees of freedom in the system.
The degeneracy of the proposed non-Kramers doublet should
also be lifted with application of a uniaxial strain; again
the signatory entropy associated with the two-channel Kondo
model should be observed.

\item The crystal field schemes proposed by Amitsuka et al.\cite{amitsuka93}
and by Santini\cite{santini98} are for
the dilute and the dense limits, respectively.
Qualitatively they are very different; more specifically
the lowest lying state is a doublet in the scheme
of Amitsuka et al.\cite{amitsuka93} whereas it is a
singlet in Santini's proposed scenario.\cite{santini98}
If indeed there is such a dramatic shifting of the
crystal-field levels as a function of uranium density
it should lead to observable nonlinearities in the
lattice parameters and dramatic changes in the nonlinear
susceptibility\cite{ramirez94,aliev95}
as a function of uranium doping.
By contrast if the lattice parameters grow monotonically
with doping levels, we can conclude that the single-ion
physics of the dilute system and the lattice are qualitatively
similar.

\item If the underlying physics of the dense system involves
a non-Kramer's magnetic doublet, then we expect that 
a uniaxial strain and magnetic field will split this
doublet in precisely the same way, up to a scale constant
that can be deduced from the dilute limit. 
In this situation, the phase
diagram as a function of uniaxial strain will look {\sl identical}
to the phase diagram as a function of field. 
This is the definitive 
test of whether a non-Kramer's magnetic doublet underpins
the physics of the dense lattice.  

\end{itemize}

Summarizing the discussion so far, we have presented some simple
experimental probes of time-reversal violation and the local
single-ion physics that, if observed, will substantially advance
our basic understanding of the underlying order in \uru.  We should
now
look ahead to the constraints that our discussion
imposes on any future microscopic theories in \uru. 
Such theories must provide :
\begin{itemize}

\item A description of the local single-ion physics
that is consistent with the heavy fermion behavior.

\item A description of how the hidden order emerges
from the local ion physics. Clearly, the character of the
theory  depends critically on an experimental test
of whether the primary order breaks time reversal symmetry. 

\end{itemize}
It is
important to remember in this discussion 
that $URu_2Si_2$
is a heavy fermion compound, both before and after the hidden order
develops.  In the low temperature phase, the size of $
C_V/T
\sim 90  mJmol^{-1}K^{-1}$ puts this material into the category
of intermediate heavy fermion behavior.  The superconducting
transition at 1.7K also has a large specific heat anomaly
characteristic of heavy fermion superconductivity.
The dramatic contrast with 
$ThRu_2Si_2$, which is a normal, low-mass metal 
serves to emphasize that it is the local f-electron physics
of the Uranium atom which drives the unusual properties
in \uru. The large values of $\gamma$
derive from the quenching of the local ionic degrees of freedom.
Any microscopic theory of the hidden order in $URu_2Si_2$ must
respect these essential observations.

At present, the only scenario which addresses the second
item in this list is the quadrupolar
scenario of Santini.\cite{santini98} However if the 
single-ion physics of the Uranium in \uru
is described by a non-degenerate singlet state which
mixes with higher-lying singlets to produce a quadrupole,
it is very difficult  
to see how this picture can provide
the necessary degrees of freedom for the
heavy electron behavior below the transition.

By contrast let us suppose that the local-ion physics suggested by Koga
et al\cite{koga96} for dilute $U$ in $ThRu_2Si_2$ persists to stoichiometric
\uru, involving a magnetic non-Kramers doublet.
Such a
state has the capacity to provide the required low-lying 
degrees of freedom for heavy fermion behavior, but now
we must address the second point above.
One of the interesting questions here  is how the
the two-channel physics of the single ion might play 
a  role in the hidden order.  In the dense lattice,
there is the possibility 
of constructive interference 
between the Kondo effect in the two channels that
hypothetically couple to each uranium ion. This has
the potential to produce composite orbital order
that breaks time-reversal symmetry. 
If $c_1$ and $c_2$
create electrons in the two local scattering channels
coupled to the local moment, and if ${\bf S}$ is the spin
degree of freedom associated with the magnetic non-Kramer's
doublet, then such a state would involve an order parameter
of the form\cite{coleman98}
\bea
\psi(x) = -i\langle \bigl(c^{\dagger}
_1 \pmb{$\sigma$} c_2
-
c^{\dagger}_2 \pmb{$\sigma$} c_1\bigr)\cdot {\bf S}(x) \rangle
\eea
This order parameter involves hidden orbital order, and
combines aspects of the Kondo effect and orbital
magnetism into one quantity. Similar order
parameters have been proposed 
for heavy fermon superconductivity.\cite{coleman98,balatsky,bonca,poilblanc}
Should experiments confirm the equivalence of field and uniaxial
strain on the primary order parameter, then this would seem to be
an interesting  candidate for the underlying hidden order.

In conclusion, we have contrasted two classes of theory
for the hidden order in $URu_2Si_2$ and  have proposed 
measurements designed to test (i) whether the order breaks
time-reversal symmetry and (ii) whether the local physics
is described by a non-Kramers magnetic doublet.
The  results of these experiments would considerably
further our understanding of this  fascinating  
heavy-fermion superconductor. 

Research for N. Shah, P. Coleman and J.A. Mydosh at Rutgers
was supported in part by the National
Science Foundation under Grants NSF DMR 96-14999. We thank 
G. Aeppli, D. Cox, A.P. Ramirez and 
A. Schofield for stimulating discussions.

\end{document}